\documentclass[12pt,english,dvips]{scrartcl}
\usepackage{babel}
\usepackage[T1]{fontenc}
\usepackage[ansinew]{inputenc}
\usepackage{amsmath}
\usepackage{amsfonts}
\usepackage{amssymb}
\usepackage{graphicx}
\usepackage{pstricks}
\usepackage[hang]{caption2}
\usepackage{float}

\begin{document}

\begin{titlepage} \vspace{0.2in} 

\begin{center} {\LARGE \bf 

Dark Energy as a Relic of the Vacuum-Energy Cancellation?}\\ 
\vspace*{1cm}
Orchidea Maria Lecian$^{12a}$ and Giovanni Montani$^{1234b}$, \\
\vspace{0.5cm}\hfill\\

$^{1}$ICRA --- International Center for Relativistic Astrophysics.\\
$^{2}$Dipartimento di Fisica, Universit\`a di Roma ``La Sapienza'', P.le
Aldo Moro 5,\\
00185 Roma, Italy.\\
$^{3}$ENEA C.R. Frascati (Dipartimento F.P.N.), Via Enrico Fermi 45,\\
00044 Frascati, Roma, Italy.\\
$^{4}$ICRANET C. C. Pescara, \\ Piazzale della Repubblica, 10, 65100 Pescara, 
Italy.\\
$^{a}$E-mail: lecian@icra.it\\
$^{b}$E-mail: montani@icra.it
 
\vspace*{1.8cm}

\vspace*{1cm}

{\bf   Abstract  \\ } \end{center} \indent
We analyze the dynamical implications of an exponential Lagrangian density for the gravitational field, as referred to an isotropic FRW Universe. Then, we discuss the features of the generalized deSitter phase, predicted by the new Friedmann equation. The existence of a consistent deSitter solution arises only if the ratio between the vacuum-energy density and that associated with the fundamental length of the theory acquires a tantalizing negative character. This choice allows us to explain the present universe dark energy as a relic of the vacuum-energy cancellation due to the cosmological constant intrinsically contained in our scheme. The corresponding scalar-tensor description of the model is addressed too, and the behavior of the scalar field is analyzed for both negative and positive values of the cosmological term. In the first case, the Friedmann equation is studied both in vacuum and in presence of external matter, while, in the second case, the quantum regime is approached in the framework of ''repulsive'' properties of the gravitational interaction, as described in recent issues in Loop Quantum Cosmology. In particular, in the vacuum case, we find a pure non-Einsteinian effect, according to which a negative cosmological constant provides an accelerating deSitter dynamics, in the region where the series expansion of the exponential term does not hold.  
\end{titlepage}

\section*{Introduction}

One of the most puzzling questions, which has come out from the modern understanding of the Universe evolution, is certainly the present value of the cosmological constant \cite{erty}. In fact, the observations of the recession of SNIA, Super-Novae I A, (treated as standard candles) provide convincing indications for an accelerating Universe \cite{oiuy}. This surprising behavior is guaranteed by a negative-pressure contribution, and the determination of the precise equation of state for the matter that is accelerating the Universe is the present challenge for cosmologists. However, data from the cosmic-microwave background suggest that the so-called Dark Energy has reliably the features of a cosmological constant, which corresponds to about 70 percent of the critical density of the Universe. Such an amount of the cosmological term is relevant for the actual dynamics, but extremely smaller than the vacuum value. Estimations of the vacuum energy yield indeed the Planckian value, corresponding to $10^{120}$ times the observed numbers. This striking contradiction between the theoretical predictions and the actual value suggest that, if the Universe acceleration is really due to a cosmological constant, then a precise mechanism of cancellation must be fixed for the vacuum energy density. We stress that no fundamental theory provides a convincing explanation for such a cancellation and therefore it is naturally expected to find it from specific features of the field dynamics. The main interesting proposals to interpret the presence of Dark Energy can be divided into two classes \cite{okm}: those theories that make explicitly presence of matter and the other ones, which relay on modifications of the Friedmann dynamics. Here, we address a mixture of these two points of view, with the aim of clarifying how the ''non-gravitational'' vacuum energy affects so weakly the present Universe dynamics \cite{wei}. In what follows, we determine the Friedmann equation corresponding to an exponential form for the gravitational-field Lagrangian density. The peculiar feature of our model is that the geometrical components contain a cosmological term too, whose existence can be recognized as soon as we expand the exponential form in Taylor series of its argument. An important feature of our model arises when taking a Planckian value for the fundamental parameter of the theory (as requested by the cancellation of the vacuum-energy density). In fact, as far as the Universe leaves the Planckian era and its curvature has a caracteristic lenght much greater than the Planckian one, then the corresponding exponential Lagrangian is expandible in series, reproducing General Relativity (GR) to a high degree of approximation. As a consequence of this natural Einsteinian limit (which is reached in the early history of the Universe), most of the thermal history of the Universe is unaffected by the generalized theory. The only late-time effect of the generalized framework consists of the relic cosmological term actually accelerating the Universe. Indeed, our model is not aimed at showing that the present Universe acceleration is a consequence of non-Einsteinian dynamics of the gravitational field, but at outlining how it can be recognized from a vacuum-energy cancellation. Such a cancellation must take place in order to deal with an expandable Lagrangian term and must concern the vaccum-energy density as far as we build up the geometrical action only by means of fundamental units. The really surprising issue fixed by our analysis is that the deSitter solution exists in presence of matter only for a negative ratio between the vacuum-energy density and the intrinsic cosmological term, $\epsilon_{vac}/\epsilon_{\Lambda}$. We can take the choice of a negative value of the intrinsic cosmological constant, which predicts an accelerating deSitter dynamics. Nevertheless, in this case, we would get a vacuum-energy density greater than the modulus of the intrinsic term. This fact looks like a fine-tuning, especially if we take, as we will do below, a Planckian cosmological constant. The vacuum-energy density is expected to be smaller than the Planckian one by a factor $\mathcal{O}(1)\times\alpha^{4}$, where $\alpha<1$ is a parameter appearing in non-commutative formulations of the relativistic particle. \\
The analysis of the corresponding scalar-tensor model helps us shed light on the physical meaning of the sign of the cosmological term. In fact, for negative values of the cosmological term, the potential of the scalar field exhibits a minimum, around which scalar-field equations can be linearized. The study of the deSitter regime shows that a comparison with the modified-gravity description is possible in an off-shell region, i.e., in a region where the classical equivalence between the two formulations is not fulfilled. However, despite this apparently unphysical character, this choice is allowed by recent developments in Loop Quantum Cosmology (LQC). In fact, Ashtekar et al \cite{asht} have shown that, over a critical value, fixed by the Immirzi parameter \cite{immi} $\gamma$, the effective cosmological energy density becomes negative, and gravity exhibits a repulsive character, so that a Big Bounce of the Universe is inferred. The reason why our scheme can solve the problem of the cosmological constant is that our generalized Friedmann equation not only acquires the negative ratio discussed above, but also states that its value is $(-1+x)$, where $x$ is less/equal to the squared ratio between the Planckian length an the present Hubble radius.\\
We can summarize our point of view by the assumption that we want to build up a generalized gravitational action, which depends on the Planck length as the only parameter. This statement leads naturally to the exponential form of the Lagrangian density, hence it provides the framework of our proposal. By other words, the cancellation that takes place between the intrinsic term and the effective vacuum energy leaves a relic term, of order $10^{-120}$ times the present Universe Dark Energy, much smaller than the original.\\ 
It is worth stressing that this semi-cancellation could be treated in the usual Einstein-Hilbert (EH) scheme by introducing a positive cosmological term which compensates for the vacuum-energy density. The Einsteinian regime of the exponential Lagrangian density can be recovered after a series expansion. Nevertheless, two different possibilities are found: the series expansion either does not hold or brings puzzling predictions about the cosmological term. Correspondingly, in the first case, an unlikely implication would appear when dealing with a non-Einsteinian physics on all astrophysical scales, and, in the second case, the expansion is only possible in the region $\Lambda>>R$, i.e., in the region where the cosmological constant dominates the dynamics, but for the fact that $R$ should be the same order of $\Lambda$. This contradiction can only be solved if a suitable cancellation mechanisms is hypothesized: here we find the constraint on the ratio $\epsilon_{vac}/\epsilon_{\Lambda}\ll1$ in the deSitter regime in presence of matter.\\
In the first Section, we will analyze the problem of generalized Einstein equations from an extension of the EH action within a metric approach. In fact, for a generic function $f(R)$, field equations will be evaluated after variation with respect to the metric tensor; in the limit $f(R)\rightarrow R$, standard Einstein equations will be recovered. If a generic metric in the Jordan frame is considered, i.e. $ds^{2}=N^{2}dt^{2}-a^{2}dl^{2}$, the function $f(R)$ will depend on $a$ and on $N$: these two functions are the Lagrangian variables with respect to which the total action will be varied. These two variations will lead to the $00$ and $ii$ components of the Einstein equations, respectively, and the equivalence will be explained in the synchronous case.\\
The corresponding scalar-tensor model is briefly reviewed in Section 2, where the equivalence between $f(R)$ theories and a scalar field in General Relativity is discussed.\\
In Section 3, the interpretative problems of the vacuum energy will be introduced, and the necessity to establish a cut off will be approached, as an example, within the formalism of the modified canonical commutation relations predicted by the generalized uncertainty principle (GUP).\\ 
The fourth Section is aimed at investigating the features of an exponential gravitational action, as far as the deSitter regime is concerned, and the appearance of a negative energy density will be regarded to as a means to explain the present small value of the Universe vacuum energy.\\
The scalar-tensor model for an exponential Lagrangian density will be analyzed in Section 5, and, for the deSitter phase, the scalar field will be shown to admit a damped oscillating solution that tends to the fixed (minimum) value.\\
A proposal for the solution of the puzzle is eventually exposed in Section 6, where the Universe acceleration is related to the vacuum energy through the introduction of the dimensionless parameter $\delta$, which acts like a compensating factor between the energy density associated to the cosmological constant and that estimated for the vacuum energy in presence of a cut-off. \\
In Section 7, concluding remarks follow.\\

\section{Generalized Gravitational Action}
The dynamics of a gravitational field associated to the metric tensor $g_{\mu\nu}$, $\mu=0,1,2,3$, coupled to matter is described, as usual, by the following total action:
\begin{equation}
S=S_{EH}+S_{M},
\end{equation}
where $S_{EH}$ denotes the EH action, and reads
\begin{equation}
S_{EH}=-\frac{c^{3}}{16\pi G}\int d^{4}x \sqrt{-g} R,
\end{equation}
while $S_{M}$ refers to the matter contribution, and can be expressed via the sum over all the matter fields $L_{f}$, as
\begin{equation}
S_{M}=\frac{1}{c}\sum _{f}\int d^{4}x\sqrt{-g} L_{f},
\end{equation}
and $\sqrt{-g}\equiv det g_{\mu\nu}$. The form of the gravitational action is fixed by the request that the field equations contain second-order derivatives of the metric tensor only, and, by its variation with respect to the metric tensor $g_{\mu\nu}$, it is natural to recognize the well-known Einstein equations
\begin{equation}
R_{\mu\nu}-\frac{1}{2}g_{\mu\nu}R=\frac{8\pi G}{c^{4}}T_{\mu\nu},
\end{equation}
where $T_{\mu\nu}$ is the stress-energy tensor.\\
In the application of the variational principle, the natural assumption that $\delta g_{\mu\nu}$ vanishes on the boundaries of the considered space-time region has to be adopted unless boundary terms are included into the original action. We also note that we could get second-order equations from $S_{EH}$ because of the appearance of surface contribution to the Ricci scalar. This feature does not hold when, instead of $R$, we take a generic function of it; such a more general choice is yet compatible with the 4-diffeomorphism invariance of the theory, but it could lead to very different dynamical implications for the gravitational field. Over the years, many proposals have been addressed in this direction \cite{a1}, based on the idea that higher-order contributions could become important where the space-time curvature takes very large values, and could remove unphysical singularities from the theory \cite{a2}.\\ 
We now fix our attention to the generalization of the previous scheme when the following gravitational action is taken into account
\begin{equation}\label{fdr}
S_{G}=-\frac{c^{3}}{16\pi G}\int d^{4}x \sqrt{-g} f(R)
\end{equation}
whose variation with respect to $g^{\mu\nu}$ yields the generalized Einstein equations
\begin{equation}
-\frac{1}{2}g_{\mu\nu}f(R)+f'(R)R_{\mu\nu}-\nabla_{\nu}\nabla_{\nu}f'(R)+g_{\mu\nu}\nabla_{\rho}\nabla^{\rho}f'(R)=\frac{8\pi G}{c^{4}}T_{\mu\nu},
\end{equation}
where $f'(R)\equiv df(R)/dR$.\\
It is immediate to recognize that, for $f(R)=R$, the usual Einstein dynamics is recovered. As expected, the new field equations contain higher-order derivatives, and, in particular, forth-order derivatives of the metric tensor appear.

\subsection{Lagrangian approach for the FRW model}
As an application of the generalized gravitational theory discussed in the previous section, let's now consider the following FRW line element
\begin{equation}
ds^{2}=N(t)^{2}dt^{2}-a(t)^{2}dl^{2},
\end{equation}
$N(t)$ being the lapse function, $a(t)$ the cosmic scale factor of the Universe and $dl^{2}$ reading 
\begin{equation}
dl^{2}=\frac{dr^{2}}{1-kr^{2}}+r^{2}\left( d\theta ^{2}+\sin ^{2} \theta d\phi ^{2} \right),
\end{equation}
with $0<r<1$, $0\le \theta <\pi$, $0\le \phi <2\pi$ and $k=0, \pm 1$ denoting the sign of the spatial curvature \cite{kotu}.\\
Taking into account the homogeneity request (which implies the energy density $\epsilon=\epsilon(t)$ and the pressure $p=p(t)$) and applying to a fixed volume of the expanding Universe the first thermodynamical principle, $dU=\delta Q-pdV$, (with the isoenthropic character of the Universe, $\delta Q=0$), we provide the following relation
\begin{equation}\label{28}
d(\epsilon a^{3})=-3pa^{2}da.
\end{equation}
In view of the homogeneity of the space-time, the action for the FRW model reduces to a 2-dimensional problem: in fact, the action for the cosmic scale factor $a(t)$ and the lapse function $N(t)$ reads
\begin{equation}
S= -\frac{Vc^{4}}{16G\pi}\int dt Na^{3} f(R)-V\int dt Na^{3}\epsilon (a),
\end{equation}
where $V$ is the volume of the space portion on which the action is taken.\\
So far, varying this action with respect to $N$, we obtain
\begin{equation}
a^{3}f+Na^{3} f'\frac{\partial R}{\partial N}+\frac{16G\pi}{c^{4}}\epsilon-6f''\frac{dR}{dt}\frac{a^{2}\dot{a}}{N^{2}}-6f'\left[\frac{a^{2}\ddot{a}}{N^{2}}+2\frac{a\dot{a}^{2}}{N^{2}}-2\frac{a^{2}\dot{a}\dot{N}}{N^{3}}\right]=0,
\end{equation}
and, in the synchronous reference, $N=1$, 
\begin{equation}\label{one}
\frac{1}{2}f+3f'\frac{\ddot{a}}{a}-3f''\frac{dR}{dt}\frac{\dot{a}}{a}=-\frac{8G\pi}{c^{4}}\epsilon ,
\end{equation}
which is the same as the $00$-component of the generalized Einstein equations for the FRW metric, and reduces to the standard Friedmann equation when $f(R)\equiv R$.\\
On the other hand, variation with respect to $a$ leads to the generalized Euler-Lagrange equation
\begin{equation}\label{29}
\frac{\partial L}{\partial a}-\frac{d}{dt}\frac{\partial L}{\partial \dot{a}}+\frac{d^{2}}{dt^{2}}\frac{\partial L}{\partial \ddot{a}}=0.
\end{equation}
The validity of this equation requires that $\delta \dot{a}$ vanish on the boundaries; nevertheless, a large class of variation functions is still available for the calculation.
Expressing (\ref{29}) for $N=1$, we find
\begin{equation}\label{two}
-\frac{1}{2}f+f'\left[-\frac{\ddot{a}}{a}-2\frac{\dot{a}^{2}}{a^{2}}-2\frac{k}{a^{2}}\right]+2f''\frac{dR}{dt}\frac{\dot{a}}{a}+f'''(\frac{dR}{dt})^{2}+f''\frac{d^{2}R}{dt^{2}}=-\frac{8G\pi}{c^{4}}p
\end{equation}
which coincides with the $ii$-components of the generalized Einstein equations in the FRW metric. We stress that the equation above has been obtained making use of the continuity equation (\ref{28}).\\
Equations (\ref{one}) and (\ref{two}) describe the whole dynamics of the FRW Universe in a synchronous reference frame, when the gravitational Lagrangian is generalized, as in (\ref{28}).\\
Finally, combining together (\ref{one}) and (\ref{two}), we can restate, for our general case, the well known equation for the universe acceleration  
\begin{equation}
\frac{1}{6}\left( f'''(\frac{dR}{dt})^{2}+3f''\left[\frac{d^{2}R}{dt^{2}}+\frac{\dot{a}}{a}\frac{dR}{dt}\right]+f'(-2\frac{\dot{a}^{2}}{a^{2}}-2\frac{k}{a^{2}})-f\right)=-\frac{4\pi G}{3c^{4}}(\epsilon +3p).
\end{equation}
The results obtained in this section are at the ground of our cosmological investigation based on the generalized gravitational action \cite{883}.\\
To conclude, we note that (\ref{one}), (\ref{two}) and (\ref{28}) are among them correlated ; in fact, as it can be easily checked after straightforward calculation, differentiating (\ref{one}) and using the continuity equation (\ref{28}) , we generate (\ref{two}). Thus, as in the standard case, here we deal with the three unknowns $\epsilon$, $p$ and $a$ and two independent equations only arise for them: as a consequence, to develop a solution of our generalized FRW dynamics, the equation of state $p=(\gamma -1)\epsilon$ is required, i.e.,
\begin{equation}
\epsilon(a)=\mathcal{C}a^{-3\gamma},\mathcal{C}=const.,
\end{equation}  
\begin{equation}
p=(\gamma-1)\mathcal{C}a^{-3\gamma},
\end{equation}
$\gamma$ being the polytropic index.

\section{Scalar-tensor gravity}
It is possible to demonstrate that the non-linear gravitational Lagrangian (\ref{fdr}) can be cast in a dynamically-equivalent form, i.e., the action for a scalar field in GR (with a rescaled metric), by means of a suitable conformal transformation. Similarly, in Brans-Dycke theory, and, in general, in scalar-tensor theories, the original variables describing the metric and a scalar field in the Jordan frame can be transformed into two new variables describing the scalar field minimally coupled to gravity, in the Einstein frame. The two descriptions are extremely interconnected \cite{soti}, and can be interpreted as two different mathematical versions of the same physics \cite{rob07}.\\  
In order to illustrate this equivalence \cite{noji03}, as a first step, two auxiliary fields, which play the role of Lagrange multipliers, can be introduced, and then, as a second step, a suitable conformal transformation for the metric tensor is performed, in order to bring the action in the usual form.\\
The first step consists in introducing the two Lagrange multipliers $A$ and $B$, which allow one to rewrite (\ref{fdr}) as
\begin{equation}\label{mult}
S=\frac{1}{k^{2}}\int d^{4}x\sqrt{-g}\left[B(R-A)+f(A)\right],
\end{equation}
where variation with respect to $B$ leads to $R=A$, while variation with respect to $A$ gives the identities
\begin{equation}
B=f'(A),
\end{equation}
or, equivalently,
\begin{equation}
A=g(B).
\end{equation}
It is possible to eliminate either $A$ or $B$ from (\ref{mult}), thus obtaining
\begin{equation}\label{subst}
S=\frac{1}{k^{2}}\int d^{4}x\sqrt{-g}\left[B(R-g(B))+f(g(B))\right]
\end{equation}
or
\begin{equation}\label{elimination}
S=\frac{1}{k^{2}}\int d^{4}x\sqrt{-g}\left[f'(A)(R-A)+f(A)\right],
\end{equation}
respectively. Equations (\ref{subst}) or (\ref{elimination}) are equivalent, at least from a classical point of view, and are usually referred to as the Jordan-frame action in presence of the two auxiliary fields.\\
Furthermore, the conformal scaling of the metric tensor  
\begin{equation}\label{conf}
g_{\mu\nu}\rightarrow e^{\phi}g_{\mu\nu}
\end{equation}
can be taken into account, in order to put the previous results in the Einstein frame. For the particular choice $\phi=-\ln f'(A)$, action (\ref{elimination}) reads
\begin{equation}\label{scaltens}
s=\frac{1}{k^{2}}\int d^{4}x\sqrt{-g}\left[R-\frac{3}{2}g^{\rho\sigma}\partial_{\rho}\phi\partial_{\sigma}\phi-V(\phi)\right],
\end{equation}
where 
\begin{equation}
V(\phi)=\frac{A}{f'(A)}-\frac{f(A)}{f'(A)^{2}},
\end{equation}
i.e., this action describes a scalar field minimally-coupled to the rescaled metric.\\
So far, the non-linear modified gravitational action (\ref{fdr}) and the scalar-tensor model (\ref{scaltens}) are very deeply linked, and can be interpreted, on shell (i.e., for $\phi=-\ln f'(A)$), as two different models describing the same physics.\\
If a matter fluid is taken into account, the pertinent stress-energy tensor $T_{\mu\nu}$ associated to the energy density $\epsilon$, the pressure $p$ and the four-velocity $u_{\mu}$, $T_{\mu\nu}=(\epsilon+p)u_{\mu}u_{\nu}-pg_{\mu\nu}$, has to be rescaled as 
\begin{equation}\label{conf2}
T_{\mu\nu}\rightarrow e^{-\phi}\hat{T}_{\mu\nu}, \quad T^{\mu}_{\quad\nu}\rightarrow e^{-2\phi}\hat{T}^{\mu}_{\quad\nu},
\end{equation}
according to the conformal transformations induced by (\ref{conf}), i.e.,
\begin{equation}\label{conf3}
u_{\mu}\rightarrow e^{\phi/2}\hat{u}_{\mu}, \quad \epsilon\rightarrow e^{-2\phi}\hat{\epsilon}, \quad p\rightarrow e^{-2\phi}\hat{p}.
\end{equation}
Nevertheless, the interpretation of the equivalence between the two models gives rise to some remarks about the physical meaning of the transformation \cite{magn93}. In the Jordan frame, gravity is described by the metric tensor only, while, in the Einstein frame, the rescaled metric tensor experiences the scalar field as a source matter field. These considerations entail the discussion of the role of matter fields coupled to gravity. In fact, non-linear theories of gravity in vacuum \cite{vac96} leave room for ambiguity about which frame should be considered as the physical one, while the presence of matter fields sheds light on this indistinctness by the request of a minimal coupling with gravity. The mathematical equivalence between the two theories is achieved dynamically, since the spaces of the classical (on shell) solutions are locally isomorphic.\\
Without aiming at solving this interpretative ambiguity \cite{far99}, throughout this paper we will try to investigate the role and the properties of matter field \cite{kiov} in the determination of cosmological solutions \cite{cat06}.

\section{The vacuum-energy problem}
As well known \cite{a3}, the vacuum-energy density associated to a massless quantum field is a diverging quantity unless an appropriate normal ordering (which, on curved space-time, would depend on the metric properties of the manifold) can be found; however, if we fix a cut-off on the momentum variable, $P_{max}=\alpha \frac{\hbar}{l_{pl}}$ ($\alpha$ being a dimensionless parameter of order unity) then the vacuum energy density can be estimated as follows
 \begin{equation}\label{ved}
\epsilon_{vac}=\int_{0}^{P_{max}} \frac{d^{3}p}{\hbar^{3}}cp=\int_{0}^{P_{max}} \frac{4\pi p^{2}dp}{\hbar^{3}}cp=\pi\alpha^{4}\epsilon_{pl},
 \end{equation}
where $\epsilon_{pl}\equiv\hbar c/l_{pl}^{4}$.\\  
By other words, we would have to deal with a vacuum-energy density of a Planck-mass particle per Planck volume. A more rigorous understanding for the parameter $\alpha$ comes out from an approach based on GUP. Such theories implement modified canonical operators obeying the generalized relation
\begin{equation}
[x,p]=i\hbar(1+\frac{1}{\alpha^{2}}\frac{G}{c^{3}\hbar}p^{2}).
\end{equation}
This commutation relation can be recognized on the ground of fundamental properties of the Minkowski space in presence of a cut-off, but it also comes out from quantum-gravity and string-theory approaches \cite{mnbv}. As a consequence of non-commutative models, we deal with a notion of minimal length associated to a particle state. For instance, in the case of a non-relativistic particle \cite{lkjh}, we get the following limit for its wave-length
\begin{equation}
\lim_{E\rightarrow\infty}\lambda(E)=\frac{4}{\alpha} l_{pl},
\end{equation}
$E$ being the energy of the particle.\\
For a discussion of a maximum value for a relativistic-particle momentum at Planck scales, in the context of the k-Poincar\'e algebra, see \cite{sdfg}. But it is worth noting that, in our case, the discussion above must be referred to a flat FRW background, and, therefore, all the observables correspond with physical quantities corrected by the presence of the scale factor.\\   
However, no evidence appears today for such a huge cosmological term; in fact, recent observations on Supernova data \cite{oiuy} indicate that the Universe is now accelerating with a non-definite equation of state \cite{asdf}. The indication from cosmic microwave background anisotropies suggests one that the most appropriate characterization of such an equation of state be $p\sim -\epsilon$ \cite{xcvb}\cite{erty}. However, when estimating this observed cosmological term, it is immediate to recognize that it is extremely smaller than the cut-off value. In fact, for the observed value of the constant energy density, we get the estimation
\begin{equation}\label{poi}
\epsilon_{today}\sim 0.7\epsilon_{0}\cong \frac{2c^{2}H_{0}^{2}}{8\pi G}=\frac{1}{4\pi\alpha^{4}}\left(\frac{l_{pl}}{L}\right)^{2},
\end{equation}
where $\epsilon_{0}$ denotes the present Universe critical density $\epsilon_{0}\sim\mathcal{O}(10^{-29})g cm^{-3}$, $H_{0}\sim 70 Km s^{-1} Mpc^{-1}$ the Hubble constant, and $\epsilon_{today}$ the present value of the vacuum-energy density; since $L_{H}\equiv cH_{0}\sim\mathcal{O}(10^{27}cm^{-1})$, we see that a large factor $10^{-120}$ appears in (\ref{poi}), i.e. $\epsilon_{today}\sim\mathcal{O}(10^{-120})\epsilon_{vac}$. It is well known that this striking discrepancy between the expected and the observed value of the vacuum energy constitutes one of the greatest puzzle of modern cosmology \cite{wei}. Below, we will propose a solution to such a puzzle based on the peculiarity that a deSitter dynamics acquires in the context of an exponential form of the gravitational Lagrangian.

\section{Exponential Lagrangian Density}
Recent observations based on Supernova \cite{oiuy} data indicate that the universe is now accelerating with a non-definite equation of state \cite{asdf}. The indication from cosmic-microwave-background anisotropies suggests one that the most appropriate characterization of such an equation of state be $p\sim -\epsilon$ \cite{erty}, i.e. a cosmological term. The appearance of a non-zero cosmological constant indicates that $f(R=0)\neq 0$, while the EH action is a linear term in $R$, with the same sign of the previous one. To deal with $f(R)$ as a series expansion, we would have, in principle, to fix an infinite number of coefficients. However, in what follows, we address the point of view that only one characteristic length fixes the dynamics: the cosmological constant $\Lambda$, apart from the Planck length, which can be constructed with the fundamental units $G,c,h$ as $l_{pl}\equiv \sqrt{(G\hbar/c^{3})}$. As a consequence of this point of view, we fix the following explicit form for $f(R)$ 
\begin{equation}\label{cat5}
f(R)=\lambda e^{\mu R},
\end{equation}
where $\lambda$ and $\mu$ are two constants available for the problem. For a discussion of the local dynamical stability of a Universe described by this kind of Lagrangian, see \cite{carlos}. Comparing the first two terms that come from the expansion of (\ref{cat5}) (valid in the region $\mu R\ll 1$ ), with the EH action plus a cosmological term, i.e. 
\begin{equation}
L=-\frac{\hbar}{16 \pi l^{2}_{P}}\left( R+2\Lambda \right),
\end{equation}
we arrive at the following identifications
\begin{equation}
\lambda=2\Lambda,\ \ \mu=\frac{1}{2\Lambda}.
\end{equation}
As required, our gravitational Lagrangian is fixed by one parameter only, which has to be provided by observational data.\\
When specified for the present choice of the Lagrangian density, the two field equations for the FRW model, (\ref{one}) and (\ref{two}), take the explicit expressions
\begin{equation}\label{onemu}
\Lambda e^{\mu R}+3 e^{\mu R}\frac{\ddot{a}}{a}-\frac{3}{2\Lambda} e^{\mu R} \frac{dR}{dt}\frac{\dot{a}}{a}=-\frac{8G\pi}{c^{4}}\epsilon
\end{equation}
\small
\begin{equation}\label{twomu}
-\Lambda e^{\mu R}+ e^{\mu R}\left[-\frac{\ddot{a}}{a}-2\frac{\dot{a}^{2}}{a^{2}}-2\frac{k}{a^{2}}\right]+\frac{1}{\Lambda} e^{\mu R}\frac{dR}{dt}\frac{\dot{a}}{a}+\frac{1}{(2\Lambda)^{2}} e^{\mu R}(\frac{dR}{dt})^{2}+\frac{1}{2\Lambda} e^{\mu R}\frac{d^{2}R}{dt^{2}}=-\frac{8G\pi}{c^{4}}p,
\end{equation}
\normalsize
respectively.\\
These two equations have to be coupled to the continuity equation (\ref{28}) and to the equation of state $p=(\gamma -1)\epsilon$. 

\paragraph{deSitter regime}\null As a first step of this generalized FRW dynamics, we face the study of the deSitter model, where a constant vacuum energy density is taken into account. To this end, we take a cosmic scale factor of the form
\begin{equation}
a=a_{0} e^{\sigma t},\ \ a_{0}=const,\ \ \sigma =const.
\end{equation}
It is easy to recognize that for such a choice, the Ricci scalar rewrites $R=-12\sigma ^{2}/c^{2}$; hence, according to the equation of state of a cosmological constant $p=-\epsilon$,  equation (\ref{onemu}) reduces to the simple form  
\begin{eqnarray}\label{sopra}
\epsilon=-\epsilon_{\Lambda}e^{-x}\left(1+\frac{x}{2}\right)\\
\epsilon_{\Lambda}\equiv\frac{c^{4}\Lambda}{8\pi G},x\equiv \frac{6\sigma^{2}}{c^{2}\Lambda}\, .
\end{eqnarray}
The expression above has the surprising feature that the energy density would acquire a negative sign. This unphysical property of the model is formally removed as soon as we expand the exponential term in correspondence to small values of the dimensionless constant $x$, and, restating the usual Friedmann relation, we get 
\begin{equation}\label{qwert}
\epsilon=\epsilon_{\Lambda}\left( \frac{x}{2}-1\right)\Rightarrow \sigma^{2}=\frac{8\pi G}{3c^{2}}\left( \epsilon+\epsilon_{\Lambda}\right).
\end{equation}
Thus, when the expansion rate of the Universe $\sigma$ is much smaller then the cosmological constant $\Lambda$, we get the usual Friedmann relation between matter and geometry. But, though such a standard relation is apparently reproduced as a low-curvature approximation for $x\ll1$, nevertheless its inconsistency shows up when (\ref{qwert}) is restated as 
\begin{equation}\label{14}
x=2\left(\frac{\epsilon}{\epsilon_{\Lambda}}+1\right).
\end{equation}
We see that, by the expression above, for positive values of $\epsilon$ and $\epsilon_{\Lambda}$, the quantity $x$ has always to be greater than two, in clear contradiction with the hypothesis $x\ll1$, at the ground of the derivation of (\ref{qwert}). Though we are dealing with a surprising behavior, due to the negative ratio $\epsilon/\epsilon_{\Lambda}$, i.e., $\Lambda<0$, however this feature offers an intriguing scenario. In fact, in the next section, we will apply relation (\ref{14}) to treat the non-observability of the universe vacuum energy in connection with the present Universe acceleration.\\
We conclude this section by emphasizing that (\ref{sopra}) admits a special vacuum solution ($\epsilon\equiv 0$), which corresponds to the relation $x\equiv \frac{6\sigma^{2}}{c^{2}\Lambda}=-2$. For the choice of a negative cosmological constant, $\Lambda=-\mid\Lambda\mid$, the equation above provides
\begin{equation}\label{sigmaq}
\sigma^{2}=\frac{c^{2}\mid\Lambda\mid}{3}.
\end{equation}
Thus we see that, in vacuum, our model has the peculiar feature of predicting a deSitter evolution in correspondence to a negative $\Lambda$ value \cite{mal}. However it should be noted that, for this value of $x=-R/(2\Lambda)$, the exponential Lagrangian cannot be expanded, and we deal with the full non-perturbative regime with respect to the Einsteinian gravity. It is just the request to deal with an expandable Lagrangian that leads us to deal with the case $x\ll 1$ and to introduce an external matter field.

\section{Scalar-tensor exponential gravity}
The scalar-tensor formalism reviewed in Section 2 is here applied to the particular choice of the exponential Lagrangian density, in order to clarify the meaning of the relations found in the previous section.\\
The conformal scaling factor here reads
\begin{equation}
f(A)=\lambda e^{\mu A}, \quad f'(A)=e^{-\phi},
\end{equation}
where $A=-\phi/\mu$, and the potential rewrites
\begin{equation}\label{pot}
V(\phi)=-2\Lambda e^{\phi}(\phi+1). 
\end{equation}
According to the results of section 4, the on-shell relation between the Einstein frame and the Jordan one is recognized in the identification $A\equiv R\Rightarrow\phi\equiv-R/(2\Lambda)$.\\
Collecting all the terms together, we get the scalar tensor action
\begin{equation}\label{st3}
S=-\frac{c^{3}}{16\pi G} \int d^{4}x\sqrt{-g}\left[R+\frac{3}{2}g^{\mu\nu}(\partial_{\mu}\phi)(\partial_{\nu}\phi)-2\Lambda e^{\phi}(\phi+1)\right].
\end{equation}
Two remarks are now in order:
\begin{enumerate}
	\item for the comparison of (\ref{st3}) with the usual form of the scalar field, a transformation $\phi\rightarrow \sqrt\frac{16\pi G}{3c^{4}}\phi$ could be considered\footnote{We stress that the factor $\sqrt\frac{16\pi G}{3c^{4}}$ that arises in front of the kinetic part of the Lagrangian density is independent of the choice of the function $f$. We nevertheless perform this transformation in this section in order to keep the notation compact.};
	\item potential (\ref{pot}) is here referred to as one with $\Lambda$ retaining its own sign. For a discussion of the case $\Lambda>0$ in the quantum sector, where the potential admits no minimum, see Section 6. If one is interested in a description where a stable equilibrium is forecast for the field $\phi$, the sign of the constant $\Lambda$ should be reversed, i.e., $\Lambda\rightarrow -|\Lambda|$, as illustrated in figure 1.\\
\begin{figure}[ht]
\begin{center}
\includegraphics[width=0.5\textwidth]{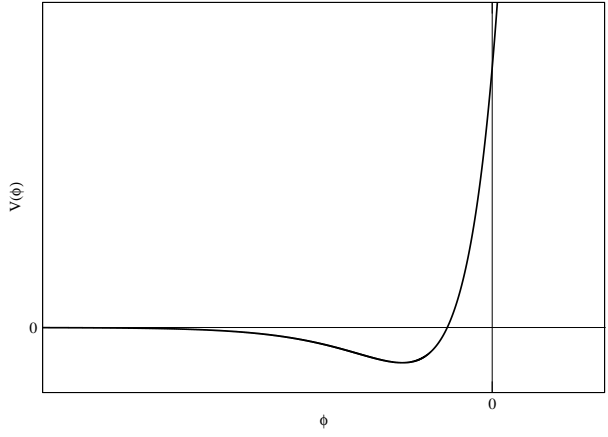}
\caption{$V(\phi)$ vs $\phi$ with $\Lambda=-\mid\Lambda\mid$ (arbitrary units).}
\end{center}
\end{figure}	
This way, the constant $\Lambda$ does not describe the observed cosmological term any more, but is a parameter of the theory, which will be tuned in order to reproduce the observational data. 
\end{enumerate}
For the considerations above, action (\ref{st3}) rewrites
\begin{equation}\label{attractive}
\begin{aligned}
S=&-\frac{c^{3}}{16\pi G} \int d^{4}x\sqrt{-g}R+\\
&+\frac{1}{c}\int d^{4}x\sqrt{-g}\left[-\frac{1}{2}g^{\mu\nu}(\partial_{\mu}\phi)(\partial_{\nu}\phi)+\epsilon_{|\Lambda|} e^{\sqrt{\frac{16\pi G}{3c^{4}}}\phi}\left(\sqrt{\frac{16\pi G}{3c^{4}}}\phi+1\right)\right],
\end{aligned}
\end{equation}
whose variations lead to the scalar-tensor Einstein equations in presence of a matter source described by the energy-momentum tensor $T_{\mu\nu}$
\begin{equation}\label{stee}
R_{\mu\nu}-\frac{1}{2}g_{\mu\nu}R=\frac{8\pi G}{c^{4}}\left(\hat{T}_{\mu\nu}e^{-\sqrt{\frac{16\pi G}{3c^{4}}}\phi}+T_{\mu\nu}(\phi)\right),
\end{equation}
\begin{equation}
g^{\mu\nu}\nabla_{\mu}\nabla_{\nu}\phi+\frac{dV}{d\phi}=0.
\end{equation}
If we specialize the rescaled metric tensor to the case of an isotropic Universe, then we deal with the FRW line element, and all the dynamical variables depend on time only. For this case, taking the matter source in the form of a perfect fluid with rescaled quantities, the Einstein-scalar system above rewrites
\begin{equation}\label{frieda}
\left(\frac{\dot{a}}{a}\right)^{2}=\frac{8\pi G}{3c^{2}}\left(e^{-2\sqrt{\frac{16\pi G}{3c^{4}}}\phi}\hat{\epsilon}(t)+\frac{1}{2}\dot{\phi}^{2}+\frac{1}{2}V(\phi)\right)
\end{equation}
\begin{equation}\label{frieda1}
2\frac{\ddot{a}}{a}+\left(\frac{\dot{a}}{a}\right)^{2}=-\frac{8\pi G}{c^{2}} \left(e^{-2\sqrt{\frac{16\pi G}{3c^{4}}}\phi}\hat{p}(t)+\frac{1}{2}V(\phi)\right)
\end{equation}
\begin{equation}
\ddot{\phi}+3H\dot{\phi}+c^{2}\frac{dV(\phi)}{d\phi}=0,
\end{equation}
where $H\equiv\dot{a}/a$.\\
Equations (\ref{frieda}) and (\ref{frieda1}), i.e., the $00$ and $ii$ components of the Einstein equations, are not independent, but linked by the rescaled continuity equation
\begin{equation}
\frac{d}{dt}\left(e^{-2\sqrt{\frac{16\pi G}{3c^{4}}}\phi}\hat{\epsilon}(t)\right)=\left(\hat{\epsilon}(t)+\hat{p}(t)\right)e^{-2\sqrt{\frac{16\pi G}{3c^{4}}}\phi}\frac{1}{a^{3}}\frac{da^{3}}{dt}.
\end{equation}
It's worth remarking that, for the present choice of the negative constant, the potential $V(\phi)$ admits now a minimum for $\phi=\phi_{0}\equiv -2\sqrt\frac{3c^{4}}{16\pi G}$, and the corresponding linearized equation reads
\begin{equation}\label{linearized}
\ddot{\phi}+3H\dot{\phi}+\frac{2}{3}c^{2}|\Lambda|e^{-2}\left(\phi+2\sqrt{\frac{3c^{4}}{16\pi G}}\right)=0.
\end{equation}
The appearance of this minimum is expected to become relevant in the dynamics of the scalar field because it is a well-known result that its total energy density follows the relation
\begin{equation}
\frac{d}{dt} \left(\frac{\dot{\phi}^{2}}{2c^{2}}+V(\phi)\right)=-3H\frac{\dot{\phi}^{2}}{c^{2}}<0,
\end{equation}
where we are assuming an expanding universe, i.e., $H>0$. In fact, starting with a given value of the energy density, sooner or later, the friction due to the universe expansion settles down the scalar field near its potential minimum \cite{dam93}.

\paragraph{deSitter regime}\null
All these results can apply to the deSitter phase, and a comparison with the issues of Section 4 can be addressed. Therefore, in what follows, we search for a solution of the linearized scalar field equation (\ref{linearized}), in correspondence with the choice $a(t)=a_{0}e^{\sigma t}$ and $\epsilon(t)\equiv\varepsilon=const$.\\
The linearized equation (\ref{linearized}) then rewrites
\begin{equation}\label{linearized2}
\ddot{\phi}+3\sigma\dot{\phi}+\frac{2}{3}c^{2}|\Lambda|e^{-2}\left(\phi+2\sqrt{\frac{3c^{4}}{16\pi G}}\right)=0,
\end{equation}
whose solution around the minimum is
\begin{equation}\label{campo}
\phi=-2\sqrt{\frac{3c^{4}}{16\pi G}}+e^{-\frac{3}{2}\sigma t}\left[C_{+}\cos\beta_{+}t +C_{-}\sin\beta_{-}t\right],
\end{equation}
where $C_{\pm}$ are two arbitrary constants, and
\begin{equation}\label{discriminant}
\beta_{\pm}\equiv\mp i\sqrt{|\Lambda|c^{2}}\sqrt{\frac{3x}{2}-\frac{8}{3}e^{-2}}:
\end{equation}
the discriminant (\ref{discriminant}) is negative for $x<0.24$, and, because of the prescription $x\ll1$, it is always negative, so that the field $\phi$ tends, as expected, to $\sqrt\frac{16\pi G}{3c^{4}}\phi_{0}=-2$. It is worth remarking that the on-shell relation provides the identification $\sqrt\frac{16\pi G}{3c^{4}}\phi_{0}=-\frac{6\sigma^{2}}{\Lambda c^{2}}=-2$.\\
Solution (\ref{campo}) can now be inserted in (\ref{frieda}): since the time derivative of the scalar field can be neglected in the vicinity of the minimum, the new Friedmann equation reads
\begin{equation}\label{sigmaquadro}
\sigma^{2}=\frac{8\pi G}{3c^{2}}\left( \hat{\epsilon} e^{4}-\hat{\epsilon}_{|\Lambda|}e^{-2}\right),
\end{equation}
and can be compared with (\ref{qwert}): because of the conformal transformations (\ref{conf3}) and (\ref{conf}), the two equations completely match. In fact, the different modifications to the two energy densities $\hat{\epsilon} e^{4}$ and $\hat{\epsilon}_{|\Lambda|}e^{-2}$ are due to the coupling of the scalar filed with matter and to the conformal transformation of the metric, respectively. Obviously, in both Jordan and Einstein frame, the metric structure remains that of a deSitter phase simply because the conformal factor $e^{-2}$ is nearly constant around the minimum. We recall that the value $\phi_{0}=-2\sqrt{\frac{3c^{4}}{16\pi G}}$ would correspond to the choice $x=-R/(2\Lambda)=-2$ in the Jordan frame defined in section 4. However, we easily check that, in the scalar-tensor theory, such a choice can no longer be a vacuum solution of the theory. In fact, in absence of matter ($\varepsilon\equiv 0$) we would deal with a negative cosmological constant as a source of the expansion rate of the universe. However, the correspondence between the Einstein and the Jordan frame takes place as far as we compare equation (\ref{sigmaquadro}) when a constant energy density is included with relation (\ref{qwert}) obtained for $x\ll 1$. Thus we are led to postulate an off-shell correspondence between the analysis developed for a deSitter space, in which the expansion rate of the universe is much smaller than the $\mid\Lambda\mid$ value, and the scalar-tensor approach near the stable configuration, as far as matter a source is included too. The off-shell correspondence provides us with a valuable tool to regard the potential defined in (\ref{attractive}) as an attractive configuration in the exponential-Lagrangian dynamics. Collecting the two points of view together, we can claim that, when dealing with an exponential Lagrangian, a deSitter phase exists, such that $\epsilon\sim\epsilon_{\mid\Lambda\mid}$ and it corresponds with general features in the space of the solution.

\section{Proposal for an explanation of the cosmological term}
Here we collect the issues of the previous sections together, in order to provide an explanation for the reason why the large value of the vacuum-energy density is today unobservable, or reduced to the actual cosmological constant $\mathcal{O}\left(10^{-120}\right)$ orders of magnitude smaller than it. We specify that our scheme does not fix the present cosmological term, but simply outlines the mechanism for a cancellation of the original cut-off term. 
\paragraph{Implementation of a coherent cosmological model}\null
In the light of the discussion above, which calls attention for a solution of the vacuum-energy problem, we are now ready to formulate our proposal for a coherent construction of our model. However, before piecing the jigsaw together, we must focus our attention on some relevant features, which arise from the previous analysis. The exponential Lagrangian is characterized, as established here, by a single parameter, according to which the expansion of the exponential term into power series holds, and which fixes the zero order term of such an expansion, i.e., a cosmological constant. The peculiar feature of this formulation consists of the following consideration. The exponential term is expandable only if $R/(2\mid\Lambda\mid)\ll1$, but this would imply that the dynamics must contain a cosmological term much greater than the Universe curvature, i.e., an inconsistency which apparently prevents us from recovering the Einstein limit. On the other hand, significant contributions from powers $R/(2\mid\Lambda\mid)\leq1$ would be predicted in a regime where the expansion of the exponential term does not hold.
We analyzed the deSitter regime in some detail because it turns out analytically treatable and very useful for the investigation of the behavior of the exponential term, since we deal with $R=const$. The main result we get in the pure geometrical (Jordan) frame is that the vacuum dynamics admits a deSitter phase in correspondence with a certain negative value of $\Lambda$ and the additional presence of matter is observable instead only if its constant energy density and $\epsilon_{\Lambda}$ have opposite signs. Furthermore, the vacuum solution lives in the non-Einsteinian region ($x\ll1$). On the contrary, in the scalar-tensor scheme, when the non-Einsteinian features are recast as matter source, we find an attractive picture in correspondence with the same (on-shell) value $x=-2$, but it turns out admissible only in presence of matter. These two different representations of the new cosmological dynamics can match only if we assume that the system evolution is always concerned with a constant matter contribution and if such a source nearly cancels the negative cosmological term, so that we fix $x\ll1$. The universal features of such a matter contribution and its constant value suggest one to identify it with the vacuum energy discussed in the previous sections. Moreover, the cancellation required to get $x\ll1$ is the natural scenario in which a relic dark energy can be recognized.\\
The reason why the cancellation proposed between the $\Lambda$ term and the vacuum energy density provides the right order of magnitude of the dark-energy contribution can be recognized in the following fact. By the structure of our model, the relic constant energy density must be a factor $\mathcal{O}(R/(2\mid\Lambda\mid))$ smaller than the dominant contribution $\mathcal{O}(\epsilon_{\Lambda})$. Thus if we take the vacuum energy density close to the Planckian value (as suggested in Section 4) then the actual ratio $R/(2\mid\Lambda\mid)$ is of order $\mathcal{O}(10^{-120})$. Such a quantity behaves like $\mathcal{O}(\frac{l_{Pl}^{2}}{L_{H}^{2}})$, where $L_{H}\sim\mathcal{O}(10^{27}cm)$ is the present Hubble radius of the universe. However, it must be remarked that such a consideration holds in the case $\epsilon_{\Lambda}$ and the vacuum energy density are the only contributions.\\  If, as below, an additional physical matter field is added, then the relic dark energy contribution is simply constrained to be less than the factor $R/(2\mid\Lambda\mid)$ of the vacuum energy. (For a recent approach which proposes a possible explanation for the value of the present Dark Energy contribution, see \cite{gurz2}).

\paragraph{Friedmann dynamics in the scalar-tensor scheme}\null 
Dividing the source energy density into the form
\begin{equation}
\epsilon_{mat}=\epsilon_{vac}+\rho(t),
\end{equation}
where $\rho(t)$ is a generic field contribution, then the Friedmann equation for the scalar-tensor scheme in proximity of the minimum $\phi_{0}$ reads
\begin{equation}\label{abo}
\left(\frac{\dot{a}}{a}\right)^{2}=\frac{8\pi G}{3c^{2}}\left( (\hat{\epsilon}_{vac}+\hat{\rho}(t) )e^{4}-\hat{\epsilon}_{|\Lambda|}e^{-2}\right).
\end{equation}
Since the compatibility of the Jordan- and the Einstein-frame approaches requires that the expansion rate of the Universe be much smaller than the corresponding parameter $\Lambda$, then we are led to account for the non-exact cancellation of the vacuum-energy density by the small parameter $\delta\ll1$ as follows.\\
If we take $\hat{\epsilon}_{vac}e^{4}=e^{-2}\hat{\epsilon}_{\mid\Lambda\mid}(1+\frac{\delta}{2})=\pi e^{4}\alpha^{4}\epsilon_{pl}$, i.e., $\mid\Lambda\mid\sim 8\pi^{2}\alpha^{4} e^{6}/l^{2}_{pl}$, eq. (\ref{abo}) restates
\begin{equation}\label{albero}
\left(\frac{\dot{a}}{a}\right)^{2}=\frac{8\pi G}{3c^{2}}\left( \hat{\rho}(t)e^{4}-\hat{\epsilon}_{|\Lambda|}e^{-2}\frac{\delta}{2}\right).
\end{equation}
Thus, when the constant energy density dominates, we recognize $\delta=\mathcal{O}(\frac{l_{Pl}^{2}}{L_{H}^{2}})$, since now $\epsilon_{\mid\Lambda\mid}$ has a Planckian value. We note that the factor $e^{6}$ appearing in the expression of $\Lambda$ is of course present only in the scalar-tensor theory, because of the rescaling of the involved energy densities.

\paragraph{Friedmann dynamics in the f(R)-frame}\null
On the other hand, this picture can be recovered even in the original Jordan frame, as far as we observe that, for a Planckian value of $\Lambda$, the exponential Lagrangian is expandable in power series immediately after the Planckian era of the Universe. In fact, as far as we fix $\epsilon_{\Lambda}$ at Planckian scales, then, as emphasized above, we automatically get for $\delta\equiv x$ of order $\mathcal{O}(10^{-120})$. By other words, even in the Jordan frame, our model is able to explain the vacuum-energy cancellation and to determine the amplitude of the compensating factor $\delta$ simply by the assumption that the gravitational action contain the Planck length as the only fundamental parameter.\\
If we now introduce a pure matter contribution, $\epsilon_{mat}\ll \mid \epsilon_{vac} \mid$, it is easy to recognize that the standard Friedmann equation with the present cosmological constant is recovered:
\begin{equation}
H_{0}^{2}=\frac{\Lambda c^{2}}{6}\left(\frac{\epsilon_{vac}+\epsilon_{mat}}{\epsilon_{\Lambda}}+1\right)= \frac{8\pi G}{3c^{2}}\epsilon_{mat}+\frac{\delta\Lambda c^{2}}{6}.
\end{equation}
All our considerations refer here to the deSitter solution, and, therefore, $\epsilon_{mat}$ is to be regarded as constant. However, it is naturally expected that the Friedmann equation with a small cosmological term arises as low-energy curvature of this theory for any dependence on $\epsilon_{mat}$; in fact, for our choice of $\epsilon_{\Lambda}$, the Lagrangian density of the gravitational field explicitly reads
\begin{equation}
L=\frac{\hbar}{l_{pl}^{4}}\pi\alpha^{4}e^{\frac{-Rl_{pl}^{2}}{16\pi^{2}\alpha^{4}}}.
\end{equation}
From this expression for the gravitational-field Lagrangian density, we recognize that, as far as the typical length scale $\mathcal{D}\gg l_{pl}$ of the curvature ($R\sim1/\mathcal{D}^{2}$), we can address the expansion in terms of small quantity $l_{pl}^{2}/\mathcal{D}^{2}$
\begin{equation}
L\simeq \frac{\hbar}{l_{pl}^{4}}\pi\alpha^{4}-\frac{\hbar}{16\pi l_{pl}}R+\mathcal{O}\left(\frac{1}{\mathcal{D}^{2}}\right)
\end{equation}
This approximated Lagrangian density would provide for the FRW metric the following Friedmann equation
\begin{equation}\label{free}
\left(\frac{\dot{a}}{a}\right)^{2}=\frac{8\pi G}{3c^{2}}\left[\epsilon_{mat}(t)+\epsilon_{\Lambda}\left(\frac{\epsilon_{vac}}{\epsilon_{\Lambda}}+1\right) \right].
\end{equation}
Such an approximated equation, isomorphic to (\ref{albero}), allows us to reproduce all the considerations developed about the exact deSitter case. However, the analysis performed above is relevant in the Jordan frame in outlining the necessity of the constraint $\epsilon_{vac}/\epsilon_{\Lambda}\sim-1$.\\
In fact, the exact deSitter case clarified that, for positive $\Lambda$ values, this relation is the only one able to provide the consistency of the Friedmann equation according to an exponential Lagrangian density. This feature could not be recognized by an approximated analysis, as in (\ref{free}). 

\paragraph{Hints for the quantum regime of the model}\null
As discussed in Section 5, the potential of the scalar field admits a minimum if the sign of the parameter $\Lambda$ is reversed. The present paragraph, on the contrary, is aimed at investigating a model where the potential term admits no minimum, i.e., a model with an absolute maximum and a slow-rolling regime, as described in figure 2.\\
\vspace{1cm}

\begin{figure}[H]
\begin{center}
\includegraphics[width=0.5\textwidth]{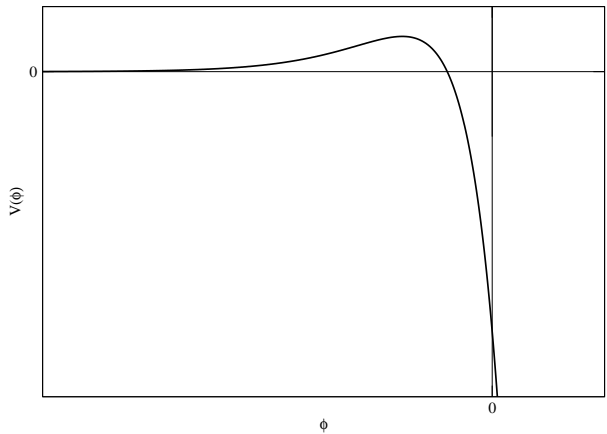}
\caption{$V(\phi)$ vs $\phi$ with $\Lambda>0$ (arbitrary units). }
\end{center}
\end{figure}
We stress that in this case the negative ratio $\epsilon_{vac}/\epsilon_{\Lambda}$ could take place only in correspondence with the apparently unphysical request $\epsilon_{vac}<0$.\\ 
A late-time solution $\phi(t)$ can be looked for, according to the potential profile, such that $\dot{\phi}(t)\rightarrow0$ in the limit $\phi(t)\rightarrow-\infty$ and $V(\phi(t)\rightarrow-\infty)\rightarrow0$. In what follows, we discuss the behavior of our scalar-tensor model near the cosmological singularity and provide some hints about its quantum dynamics. The case $\Lambda>0$, as remarked above, admits no stable configuration; the analysis below, however, would apply also in the case $\mid\Lambda\mid<0$, because the behavior of the scalar field in the vicinity of the singularity would hold as well, but for the fact that, for $\phi\rightarrow\infty$ no stable configuration would be reached. \\
In absence of external matter, i.e., $\epsilon(t)=0$, the Friedmann equation (\ref{frieda}) simplifies as
\begin{equation}
\left(\frac{\dot{a}}{a}\right)^{2}=\frac{8\pi G}{3c^{2}}\frac{\dot{\phi}^{2}}{2}:
\end{equation}
if only the positive root is taken into account, the scale factor $a$ acquires the form
\begin{equation}
a=a_{0}e^{\sqrt{\frac{4\pi G}{3c^{2}}}\phi},
\end{equation}
where $a_{0}$ is an integration constant. This way, the scale factor $a$ tends to $0$ as the field $\phi$ tends to $-\infty$.\\
As a next step, in this approximation, the linearized equation for the field $\phi$ (\ref{linearized}) reads
\begin{equation}
\ddot{\phi}+3H\dot{\phi}=0,
\end{equation}
where $H=\dot{a}/a$, so that
\begin{equation}
\dot{\phi}=\frac{\dot{\phi}_{0}}{a_{0}^{3}}e^{-\sqrt{\frac{12\pi G}{c^{2}}}\phi},
\end{equation}
$\dot{\phi}_{0}$ being an integration constant. Consequently, the time dependence of $\phi(t)$ and $a(t)$ can be found, i.e.,
\begin{equation}\label{phitronc}
\phi=\sqrt{\frac{c^{2}}{12\pi G}}\ln\left[\sqrt{\frac{12\pi G}{c^{2}}}\frac{\dot{\phi}_{0}}{a_{0}^{3}}(t-t_{0})\right],
\end{equation}
\begin{equation}\label{scalfa}
a=\frac{\dot{\phi}_{0}}{a_{0}^{2}}\sqrt{\frac{12\pi G}{c^{2}}}t^{1/3}.
\end{equation}
As requested, at the time $t_{0}=0$, the field (\ref{phitronc}) tends to $-\infty$ \cite{bel73}.\\
So far, it is possible to verify that the potential $V(\phi)$ and its first derivative could be neglected in (\ref{frieda}) and (\ref{linearized}): in fact, its contribution at early times is of order $\mathcal{O}(t^{3/2}\ln(t+1))$, which can be ignored in the presence of the leading-order terms $\mathcal{O}(t^{-2})$ due to both $(\dot{a}/a)^2$ and $\dot{\phi}^{2}$.\\
Recent studies in Loop Quantum Gravity (LQG) outlined that the expectation value of the Hamiltonian operator in a given state is, in general, different from the classical Hamiltonian contribution and this difference is responsible for systematic quantum corrections to the classical energy density involved in the problem. In particular, application of this quantum scheme to the isotropic FRW Universe (in the presence of a massless scalar field, which plays the role of time) provided modified relations between the Hubble parameter and the energy density of the Universe; this effective cosmological dynamics is mapped into the original Friedmann equation as soon as we allow the energy density of the Universe to become negative over critical values, i.e., the following correspondence takes place
\begin{equation}\label{crit}
\epsilon\rightarrow\epsilon_{eff}\equiv \epsilon\left(1-\frac{\epsilon}{\epsilon_{crit}}\right),\ \ \epsilon_{crit}=\frac{\sqrt{3}}{16\pi^{2}\gamma^{3}}\epsilon_{pl},
\end{equation}
where $\epsilon_{crit}$ (with $\gamma$ the Immirzi parameter) is a critical value of the energy density two orders below the Planck scale, over which the matter contribution becomes negative, thus illustrating a repulsive nature of the gravitational field near the (removed) cosmological singularity. In fact, in a standard Friedmann dynamics, this peculiar matter source induces a bounce in the dynamics of the scale factor, solving the singularity problem and opening interesting perspectives on the cyclic evolution of the closed Universe.\\
These developments can apply to the scalar-tensor model equivalent to the choice of an exponential gravitational action. In particular, as hinted by (\ref{phitronc}), a region can be found, where the potential $V(\phi)$ can be neglected. If external matter is absent, the results of LQG can apply to our scheme in such a region, by modifying the Friedmann equation (\ref{frieda}), i.e.,
\begin{equation}
\left(\frac{\dot{a}}{a}\right)^{2}=\frac{8\pi G}{c^{2}}\epsilon_{eff}(\phi),
\end{equation} 
where $\epsilon_{eff}(\phi)=\epsilon(\phi)\left(1-\frac{\epsilon(\phi)}{\epsilon_{crit}}\right)$, according to (\ref{crit}). In presence of external matter, on the other hand, we obtain
\begin{equation}
\left(\frac{\dot{a}}{a}\right)^{2}=\frac{8\pi G}{c^{2}}\left(\epsilon_{eff}(\phi)+\epsilon(t)e^{-2\sqrt{\frac{16\pi G}{3c^{4}}}\phi}\right).
\end{equation}
We note that we can so far analyze the implication of the dynamical equivalence between modified gravity and scalar tensor approaches. In fact, the on-shell request gives
\begin{equation}
\phi=-\mu R=-\frac{R}{2\Lambda}
\end{equation} 
and, by the power-law (\ref{scalfa}), $R=1/(3t^2)$, which does not match the solution (\ref{phitronc}) found for $\phi$, where the functional dependence on time is logarithmic. Nevertheless, as a general trend, the curvature scalar diverges as $\phi$ tends to $-\infty$,i.e., at very early times the on-shell relation is qualitatively satisfied.\\  
This analysis shows that near the cosmological singularity our scalar-tensor theory takes the form of general relativity in the presence of a massless scalar field. This fact allows us to infer some possible issues of its quantization \cite{stac}.\\
As a result, we can claim that our proposed non-Einsteinian scheme is characterized by a non-singular behavior when the corresponding scalar-tensor picture is canonically quantized. In fact, the possibility to neglect the potential field as the big-Bang is classically approached is mapped by the results discussed in \cite{asht} into a Big-Bounce. However, two relevant questions have to be faced here
\begin{enumerate}
	\item A LQG formulation for the generalized $f(R)$ gravity is not yet viable and the correspondence between the Jordan and the Einstein frame on quantum level cannot be addressed;
	\item The non-singular feature we established here in view of the possibility to neglect the potential term near the cosmological singularity can be extended to  a wide class of  scalar-tensor theories corresponding to the $f(R)$ formulation.
	\end{enumerate}
	 In particular, by the calculations above, the potential term is negligible in the asymptotic behavior towards the singularity as far as $V(\phi)$ evaluated for (\ref{phitronc}) behaves as $\mathcal{O}(t^{-2+\beta})$, with $\beta>0$. The condition on the potential term, which satisfies such a request, can be easily stated as 
\begin{equation}
\lim_{\phi\rightarrow-\infty}\frac{V(\phi)}{\phi^{\theta}e^{-2\sqrt{\frac{12\pi G}{c^{2}}}\phi}}=0,
\end{equation}
$\forall\theta>0$. Let's remark that the behaviour of a potential term $\sim e^{-2\phi}$ would correspond to a generalized gravitaitonal Lagrangian linear in the $R$ variable.

\section{Concluding Remarks}
After deriving the Einstein equations for a generalized gravitational action and specifying the results for an FRW metric, the particular choice of an exponential Lagrangian density has been analyzed.\\
The free parameters of such a Lagrangian density have been fixed as functions of the cosmological constant, and, in the deSitter regime, the ratio between the vacuum-nergy density and the geometrical contribution has been illustrated to acquire a negative sign, which has been the springboard for the investigation of the relation between the vacuum-energy and cut-off approaches to the geometrical description of the Universe. In particular, the cut-off introduced in the vacuum-energy density has been linked with the modified commutation relation following from a generalized uncertainty principle, and has been fixed at Planck scales.\\
The negative sign of the ratio $\epsilon_{vac}/\epsilon_{\Lambda}$ not only explains the non-observability of the cut-off vacuum-energy density and is in line with the LQC prediction of the Big Bounce in an FRW metric, but also allows one to recover the standard Friedmann equation in the deSitter phase, when the matter contribution is taken into account, and for any choice of the matter terms.\\
Studying some aspects of the pertinent scalar-tensor description has allowed us to investigate further connotations of the implementation of such a scheme. In particular, the physical meaning of the sign of the cosmological constant has been explained to provide interesting hints about cosmological implications. For $\Lambda<0$, the accelerating Universe is predicted to stabilize around the minimum of the scalar potential, while, for $\Lambda>0$, a possible connection with LQC has been envisaged.\\ 
The main issue of our analysis has consisted in fixing the link between the vacuum-energy cancellation and the present Universe Dark Energy. By other words, we have guessed that the actual acceleration, observed via SNIA, is due to the relic of the original huge vacuum energy, after its mean value has been compensated for by the intrinsic cosmological constant $\Lambda$ contained in the exponential Lagrangian. We will address the theoretical explanation of the phenomenologically-suggested fine tuning, $\delta\sim\mathcal{O}(10^{-120})$, as a prospective investigation.

\section{Acknowledgments}
We would like to thank Massimiliano Lattanzi for having attracted our attention on some relevant features of the generalized deSitter model.


\begin{thebibliography}{99}

\bibitem{erty}
A. Clocchiatti et al, Astrophys.J. 642 (2006) 1-21, astro-ph/0510155,\\
A.G. Riess et al., Astron.J. 116 (1998) 1009-1038, astro-ph/9805201;


\bibitem{oiuy}
R.A. Daly, S.G. Djorgovski, Astrophys.J. 597 (2003) 9-20, astro-ph/0305197,\\
G. Aldering et al., ''Supernova / Acceleration Probe: A Satellite Experiment to Study the Nature of the Dark Energy'', astro-ph/0405232;

\bibitem{okm}
S. Bludman, ''The Dynamics Driving Cosmological Acceleration: Dark Energy or Modified Gravity?'', astro-ph/0702085,\\
O. Bertolami, ''Dark Energy, Dark Matter and Gravity'', Talk presented at the International Workshop ``From Quantum to Cosmos: Fundamental Physics Research in Space'', 22-24 May 2006, Warrenton, Virginia, USA, astro-ph/0608276;

\bibitem{wei}
S. Weinberg, ''Conference Summary - 20th Texas Symposium on Relativistic Astrophysics'', astro-ph/0104482;


\bibitem{asht}
A. Ashtekar, T. Pawlowski, P. Singh, Phys.Rev.Lett. 96 (2006) 141301, gr-qc/0602086,\\
A. Ashtekar, T. Pawlowski, P. Singh, K. Vandersloot, Phys.Rev. D75 (2007) 024035, gr-qc/0612104;

\bibitem{immi}
S. Holst, Phys.Rev. D, 53 (1996) 5966-5969, gr-qc/9511026;\\

\bibitem{a1}
Y. Ezawa, H. Iwasaki, Y. Ohkuwa, N. Yamada, T. Yano, ''Correspondence between the first and the second order formalism in the generalized gravity of f(R)-type'', gr-qc/0309010,\\
S.M. Carroll et al., Phys.Rev. D71 (2005) 063513, astro-ph/0410031;\\

\bibitem{a2}
S. Nojiri, S.D. Odintsov, Mod.Phys.Lett. A19 (2004) 627-638,t hep-th/0310045,\\
I. Sawicki, W. Hu, ''Stability of Cosmological Solution in f(R) Models of Gravity'', astro-ph/0702278 ;

\bibitem{kotu}
E.W. Kolb, M.S. Turner, ''The Early Universe'' , Redwood City, California  Addison-Wesley, c1990;

\bibitem{883}
J.D. Barrow, A. C. Ottewill, J. Phys. A: Math. Gen. 16 (1983) 2757-2776;

\bibitem{soti}
T. P. Sotiriou, Class.Quant.Grav. 23 (2006) 5117-5128, gr-qc/0604028;

\bibitem{rob07} M. D. Roberts, ''Frame Dependence in Scalar-tensor Theory'', 0706.4043;

\bibitem{noji03}
S. Nojiri, S.D. Odintsov, Phys.Rev. D68 (2003) 123512, hep-th/0307288v4; 

\bibitem{magn93}
G. Magnano, L.M. Sokolowski, Phys.Rev. D50 (1994) 5039-5059, gr-qc/9312008v1;

\bibitem{vac96}
A. Hindawi, B.A. Ovrut, D. Waldram, Phys.Rev. D53 (1996) 5597-5608;

\bibitem{far99}
V. Faraoni, E. Gunzig, Int.J.Theor.Phys. 38 (1999) 217-225, astro-ph/9910176v1; 

\bibitem{kiov}
T. Koivisto, Phys.Rev. D73 (2006) 083517, astro-ph/0602031v2; 

\bibitem{cat06}
R. Catena, M. Pietroni, L. Scarabello, ''Einstein and Jordan reconciled: a frame-invariant approach to scalar-tensor cosmology'', astro-ph/0604492v1,\\
T. Clifton, J.D. Barrow, Phys. Rev. D 72 (2005) 123003; 

\bibitem{a3}
N.D. Birrell, P.C.W. Davies, ''Quantum fields in curved space'' Cambridge, Cambridge University Press, 1982-IX,\\
S. Weinberg, ''The quantum theory of fields'',Cambridge , Cambridge University press, 1998,\\ 
T. Padmanabhan, ''Why Does Gravity Ignore the Vacuum Energy?'',Contribution to the IJMPD Special Issue on Dark Matter and Dark Energy, e/print gr-qc/0609012;

\bibitem{mnbv} 
L.J. Garay, Int.J.Mod.Phys.A10 (1995) 145-166, gr-qc/9403008,\\
A. Strominger, ''Quantum Gravity and String Theory, What Have We Learned?'', Talk given at the Sixth Marcel Grossman Meeting on General Relativity, Kyoto, Japan in June 1991, hep-th/9110011;

\bibitem{lkjh}  
A. Kempf, G. Mangano, R. B. Mann, Phys.Rev. D52 (1995) 1108-1118, hep-th/9412167;

\bibitem{sdfg}  
A. Agostini, G. Amelino-Camelia, F. D'Andrea, Int.J.Mod.Phys. A19 (2004) 5187-5220, hep-th/0306013,\\
N.R. Bruno, G. Amelino-Camelia, J. Kowalski-Glikman, Phys.Lett. B522 (2001) 133-138, hep-th/0107039;\\ 

\bibitem{xcvb} 
D.N. Spergel et al., ''Wilkinson Microwave Anisotropy Probe (WMAP) Three Year Results: Implications for Cosmology'', astro-ph/0603449,\\
D.N. Spergel et al., Astrophys.J.Suppl. 148 (2003) 175, astro-ph/0302209,\\
H.V. Peiris et al., Astrophys.J.Suppl. 148 (2003) 213, astro-ph/0302225;

\bibitem{asdf}
N. Panagia, Nuovo Cim. B120 (2005) 667-676, astro-ph/0502247;

\bibitem{carlos}
 M. Abdelwahab, S. Carloni, P. K. S. Dunsby, ''Cosmological dynamics of exponential gravity'',arxiv 0706.1375;

\bibitem{mal}
J. Maldacena, Adv. Theor. Math. Phys. 2 (1998) 231, hep-th/9711200,\\
J. L. Petersen, Int.J.Mod.Phys. A14 (1999) 3597-3672, hep-th/9902131v2;

\bibitem{dam93} 
T.Damour, K.Nordtvedt, Phys. Rev. Lett. 70 (1993) 2217-2219; 

\bibitem{gurz2}
V.G. Gurzadyan, S.S. Xue, Mod.Phys.Lett. A18 (2003) 561, astro-ph/0105245,
S.G. Djorgovski, V.G. Gurzadyan, ''Dark Energy From Vacuum Fluctuations'', to appear in Proc. UCLA Conference "Dark Matter 2006", eds. D. Cline et al., Nuclear Pysics B, in press (2006), astro-ph/0610204;

\bibitem{bel73} 
V.A. Belinski, I.M. Khalatnikov, Sov. Phys. JETP, 36 (1973) 591;

\bibitem{stac}
T. Stachowiak, M. Szydlowski, ''Exact solutions in bouncing cosmology'', gr-qc/0610121.



\end{thebibliography}
\end{document}